\documentclass[prl,twocolumn,nofootinbib,superscriptaddress,showpacs]{revtex4}
\usepackage{graphicx,amsmath,amssymb,bm,multirow}

\newcommand{\be}[1]{\begin{equation}\label{#1}}
\newcommand{\ee}{\end{equation}}

\newcommand{\mev}{\, \text{MeV}}

\newcommand{\ket}[1]{|#1\rangle}
\newcommand{\braket}[2]{\langle#1|#2\rangle}
\newcommand{\matrEL}[3]{\langle#1|#2|#3\rangle}

\newcommand{\elemA}[2]{\ensuremath{{}^{#1}}\textrm{#2}}
\begin{document}

\title{{\it Ab initio} description of the exotic unbound $^7$He nucleus} 
%
%\author{Simone Baroni,$^{1,2}$ Petr Navr{\'a}til,$^{2,3}$ and Sofia Quaglioni$^3$}
%%\email[]{navratil@triumf.ca}
%\affiliation{$^1$Physique Nucl\'eaire Th\'eorique, Universit\'e Libre de Bruxelles, 
%C.P. 229, B-1050 Bruxelles, Belgium\\ $^2$TRIUMF, 4004 Wesbrook Mall, Vancouver, BC V6T 2A3, Canada\\
%$^3$Lawrence Livermore National Laboratory, P.O. Box 808, L-414, Livermore, CA 94551, USA}
%
\author{Simone Baroni}
\email[E-mail:~]{simone.baroni@ulb.ac.be}
\affiliation{Physique Nucl\'eaire Th\'eorique, Universit\'e Libre de Bruxelles, C.P. 229, B-1050 Bruxelles, Belgium}
\affiliation{TRIUMF, 4004 Wesbrook Mall, Vancouver BC, V6T 2A3, Canada}
\author{Petr Navr\'atil}
\email[E-mail:~]{navratil@triumf.ca}
\affiliation{TRIUMF, 4004 Wesbrook Mall, Vancouver BC, V6T 2A3, Canada}
\affiliation{Lawrence Livermore National Laboratory, P.O Box 808, L-414, Livermore, California 94551, USA}
\author{Sofia Quaglioni}
\email[E-mail:~]{quaglioni1@llnl.gov}
\affiliation{Lawrence Livermore National Laboratory, P.O Box 808, L-414, Livermore, California 94551, USA}
\begin{abstract}
The neutron rich exotic unbound $^7$He nucleus has been the subject of many experimental investigations. While the ground-state $3/2^-$ resonance is well established, there is a controversy concerning the excited $1/2^-$ resonance reported in some experiments as low-lying and narrow ($E_R{\sim} 1\mev$, $\Gamma{\leq 1}\mev$) while in others as very broad and located at a higher energy. This issue cannot be addressed by {\it ab initio} theoretical calculations based on traditional bound-state methods. We introduce a new unified approach to nuclear bound and continuum states based on the coupling of the no-core shell model, a bound-state technique, with the no-core shell model/resonating group method, a nuclear scattering technique. 
Our calculations describe the ground-state resonance in agreement with experiment and, at the same time, predict a broad $1/2^-$ resonance above 2 MeV.
\end{abstract}


\maketitle

%\section{Introduction}
%\label{Sec:Intro}

The $\elemA{7}{He}$ nucleus is an exotic system of three neutrons outside a $^4$He core with a particle-unstable 
$J^\pi T{=}3/2^-\, 3/2$ ground state (g.s.)
lying at $0.430(3) \mev$~\cite{Stokes_1967,Cao2012} above the threshold of a neutron and $\elemA{6}{He}$, 
which in turn is an exotic Borromean halo nucleus. 
While there is a general consensus on the $5/2^-$ resonance centered at $3.35 \mev$, 
which mainly decays to $\alpha{+}3n$~\cite{Korsheninnikov_1998}, discussions are still open for the other excited states.
In particular, the existence of a low-lying $1/2^-$ state at about $1 \mev$
has been advocated by many
experiments~\cite{Markenroth_2001,Meister_2002,Skaza_2006,Ryezayeva_2006,Lapoux_2006}
(most of them using knockout reactions with a $\elemA{8}{He}$ beam on a carbon target), 
while it was not confirmed in several others~\cite{Bohlen_2001,Rogachev_2004,Wuosmaa_2005,Wuosmaa_2008,Denby_2008,Aksuytina_2009}.
This contradictory situation arises from the main experimental difficulty 
of measuring the properties of excited states in
$\elemA{7}{He}$ in the presence of a three-body background,
coming from the particle decay of $\elemA{7}{He}$ and from the outgoing particle
involved in the reaction used to produce $\elemA{7}{He}$.
The presence of a low-lying $1/2^-$ state has also been excluded by a study on the isobaric analog states of $\elemA{7}{He}$ in $\elemA{7}{Li}$~\cite{Boutachkov_2005}. According to this latter work, a broad 
$1/2^-$ resonance at ${\sim}3.5 \mev$ with a width $\Gamma{\sim}10 \mev$ fits the data the best.
Neutron pick-up and proton-removal reactions~\cite{Wuosmaa_2005, Wuosmaa_2008} 
suggest instead a $1/2^-$ resonance at about $3 \mev$ with a width $\Gamma{\approx}2 \mev$.

The $1/2^-$ resonance controversy cannot be addressed by {\it ab initio} theoretical calculations based on traditional bound-state methods such as the Green's function Monte Carlo (GFMC)~\cite{GFMC}, the no-core shell model (NCSM)~\cite{Navratil:2000ww} or the Coupled Cluster Method (CCM)~\cite{Ha08,Hagen:2012sh,Roth:2011vt}. The complex CCM was recently applied to He isotopes, but only the g.s.\ of $^7$He was investigated~\cite{Hagen2007}.  

In this Letter, we address the low-lying resonances of $^7$He within 
the no-core shell model with continuum (NCSMC), a
new unified approach to nuclear bound and continuum states based on the coupling of the NCSM with the no-core shell model/resonating group method (NCSM/RGM)~\cite{Quaglioni:2008sm,Quaglioni:2009mn,Navratil:2010jn,Navratil2011379,Navratil:2011ay,PhysRevLett.108.042503}. In this approach,  
we augment the NCSM/RGM ansatz for the $A$-body wave function~\cite{Quaglioni:2009mn}  by means of 
an expansion over $A$-body NCSM eigenstates $\ket{A \lambda J^\pi T} $ according to: 
\begin{align}
\label{NCSMC_wav}
\!\!\!\ket{\Psi^{J^\pi T}_A} \!=\!  \sum_\lambda c_\lambda \ket{A \lambda J^\pi T} \! +\! \sum_{\nu}\!\! \int \!\! dr \, r^2 
%                               \frac{\gamma_{\nu}^{J^\pi T}(r)}{r}
                               \frac{\gamma_{\nu}(r)}{r}
                               \hat{\mathcal{A}}_\nu\ket{\Phi_{\nu r}^{J^\pi T}},
\end{align}
where the $(A-a,a)$ binary-cluster channel channel states  %$\ket{\Phi_{\nu r}^{J^\pi T}}$ contain $(A{-}a)-$ and $a-$nucleon clusters:
\begin{align}
\label{eq:formalism_20}
	\ket{\Phi_{\nu r}^{J^\pi T}} = &
	\Big[ \left(
         \ket{A-a \; \alpha_1 I_1^{\pi_1}T_1}\ket{a \; \alpha_2 I_2^{\pi_2}T_2}
         \right)^{(sT)} \nonumber\\
         &\times Y_\ell(\hat{r}_{A-a,a})
         \Big]^{(J^{\pi}T)}  \frac{\delta(r-r_{A-a,a})}{rr_{A-a,a}} \; ,
\end{align}
are labeled by the collection of quantum numbers
$\nu=\{A-a \; \alpha_1 I_1^{\pi_1}T_1; a \; \alpha_2 I_2^{\pi_2}T_2; s\ell\}$ and $\vec{r}_{A-a,a}$ is the intercluster relative vector.
The NCSM sector of the basis provides an effective description of the short- to medium-range $A$-body structure, while the NCSM/RGM cluster states make the theory able to handle the scattering physics of the system. The discrete, $c_\lambda$, and the continuous, $\gamma_{\nu} (r)$ %$\gamma^{J^\pi T}_{\nu} (r)$ 
unknowns of the NCSMC wave functions are obtained as solutions of the following coupled equations,
\begin{eqnarray}\label{eq:formalism_110}
  \left(
     \begin{array}{cc}
        H_{NCSM} & \bar{h} \\
        \bar{h}  & \overline{\mathcal{H}} 
     \end{array}
  \right)
  \left(
     \begin{array}{c}
          c \\
		  {\chi}
     \end{array}
  \right)
  = 
  E
  \left(
     \begin{array}{cc}
        1 & \bar{g} \\
        \bar{g}  & 1 
     \end{array}
  \right)
  \left(
     \begin{array}{c}
          c \\ 
		  {\chi}
     \end{array}
  \right),
\end{eqnarray}
where $\chi_{\nu} (r)$ 
are the relative wave functions in the NCSM/RGM sector when working with the orthogonalized cluster channel states~\cite{Quaglioni:2009mn}. 
The NCSM sector $H_{NCSM}$ of the Hamiltonian kernel is a diagonal matrix of the NCSM energy eigenvalues, while 
$\overline{\mathcal{H}}$ is the orthogonalized NCSM/RGM kernel~\cite{Quaglioni:2009mn}.
The coupling between the two sectors is described by the 
overlap, $\bar{g}_{\lambda \nu}(r)$, and hamiltonian, $\bar{h}_{\lambda \nu}(r)$, form factors respectively  proportional to the $\braket{A \lambda J^\pi T}{\hat{\mathcal{A}}_{\nu} \Phi_{\nu r}^{J^\pi T  }}$ and 
$\matrEL{A \lambda J^\pi T} {\hat{H} \hat{\mathcal{A}}_{\nu}} {\Phi_{\nu r}^{J^\pi T  }}$ matrix elements.
We solve the NCSMC equations by applying the coupled-channel microscopic R-matrix method on a Lagrange mesh~\cite{R-matrix}. Further details on the formalism will be given elsewhere~\cite{Baroni2012}.

We begin by presenting NCSM calculations for $^6$He and $^7$He that will serve as input for the subsequent NCSM/RGM and NCSMC investigations of $^7$He.
In this work, we use the similarity-rnormalization-group (SRG) evolved~\cite{SRG,Roth_SRG,Roth_2010,Bogner_2010} chiral N$^3$LO $NN$ potential of Refs.~\cite{Entem:2003ft,Machleidt:2011zz}. 
For the time being, we omit both induced and chiral initial three-nucleon forces, and our results depend on the low-momentum SRG 
parameter $\Lambda$. However, 
for $\Lambda = 2.02$ fm$^{-1}$, we obtain 
realistic binding energies for the lightest nuclei, e.g., 
$^4$He and, 
especially important for the present investigation, $^6$He (see Table~\ref{tab:NCSM_He_gs}).  Consequently, 
this choice of $NN$ potential allows us to perform qualitatively and quantitatively meaningful calculations for $^7$He that can be compared to experiment.
Except where 
differently stated, all results shown in this work have been obtained with an harmonic oscillator (HO) $N_{\rm max}{=}12$
basis size and frequency $\hbar\Omega{=}16$ MeV.
\begin{table}[t]
\begin{center}
\begin{ruledtabular}
\begin{tabular}{c|ccc}
 $E_{\rm g.s.}$ [MeV]           & $^4$He   &  $^6$He     &  $^7$He      \\
\hline
NCSM $N_{\rm max}{=}12$ &  -28.05   &  -28.63      &  -27.33       \\
NCSM extrap.                   & -28.22(1)& -29.25(15) &  -28.27(25) \\  
Expt.                                &  -28.30   & -29.27       &  -28.84       \\
\end{tabular} 
\end{ruledtabular}
\caption{Ground-state energies of $^{4,6,7}$He in MeV. 
An exponential fit was employed for the extrapolations. }
\label{tab:NCSM_He_gs}
\end{center}
\end{table}
%

\begin{figure}[t]
\begin{center}
\includegraphics[clip=,width=0.45\textwidth]{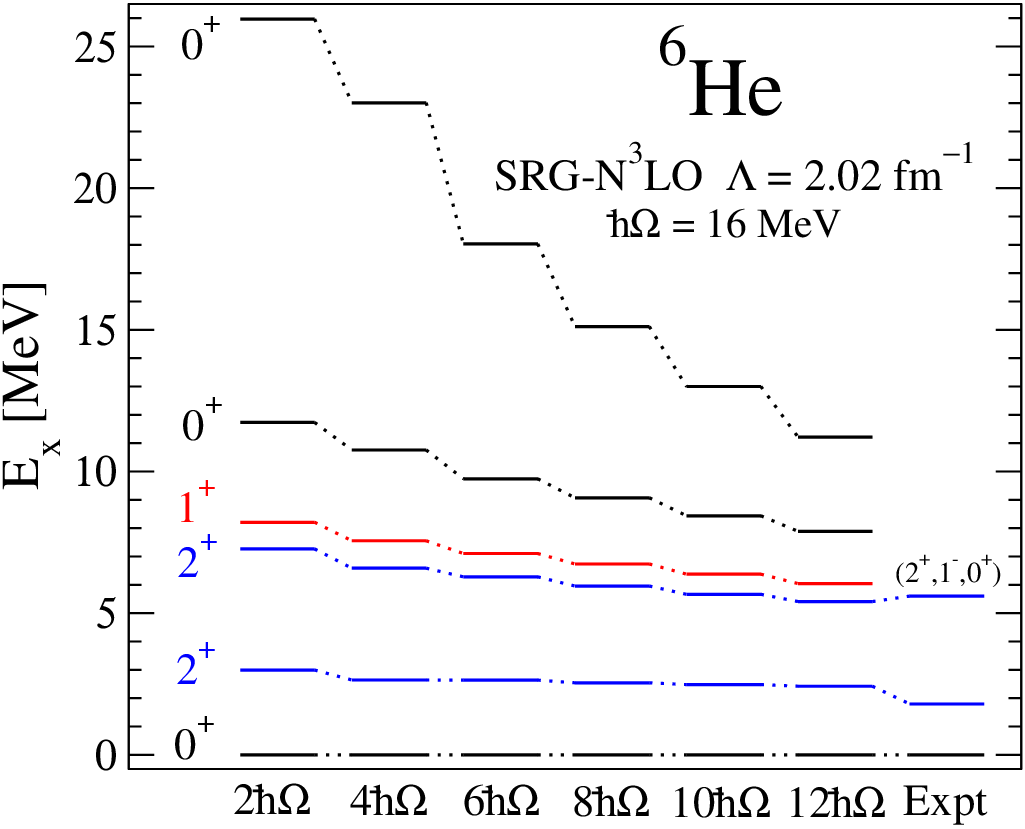}
\end{center}
\caption{(color online). Dependence of $^6$He excitation energies on the size of the basis $N_{max}$. 
%NCSM calculations were performed with the SRG-N$^3$LO $NN$ potential with $\Lambda=2.02$ fm$^{-1}$ and the HO frequency of $\hbar\Omega{=}16$ MeV.
}
\label{6He_exct}
\end{figure}

%%
%\begin{figure}[!ht]
%\begin{center}
%\includegraphics[clip=,width=0.45\textwidth]{He6srg-n3lo2.02_Egs.eps}
%\end{center}
%\caption{(color online). Ground-state energy of $^{6}$He calculated within the NCSM using the SRG-N$^3$LO $NN$ potential with $\Lambda=2.02$ fm$^{-1}$. The dependence on the HO frequency for different $N_{\rm max}$ basis sizes. The points with error bars represent results of exponential extrapolation.}
%\label{6He_gs}
%\end{figure}
%%
%Our calculated $^6$He g.s. energies for a range of HO frequencies and various basis sizes chracterized by $N_{\rm max}$ are presented in Fig.~\ref{6He_gs}. 
%
The variational NCSM calculations converge rapidly and can be easily extrapolated. 
%Extrapolated values with their error estimates and 
%$N_{\rm max}{=}12$ results for the NCSM g.s.\ energies of $^4$He, $^6$He and $^7$He  
%are given in Table~\ref{tab:NCSM_He_gs}.
At $N_{\rm max}{=}12$ (our $^{6,7}$He limit for technical reasons), the dependence of the $^6$He g.s.\ energy on the HO frequency is flat in the
range of $\hbar\Omega\sim 16{-}19$ MeV. In general, when working within an HO basis, lower frequencies are better suited for the
description of unbound systems. Therefore, we choose $\hbar\Omega{=}16$ MeV for our subsequent calculations. 
Extrapolated g.s. energies with their error estimates and the $N_{\rm max}{=}12$ results %$\hbar\Omega{=}16$ MeV 
are given in Table~\ref{tab:NCSM_He_gs}. 
Calculated $^6$He excitation energies for basis sizes up to $N_{\rm max}{=}12$ are shown in Fig.~\ref{6He_exct}.  The $^6$He is weakly bound with all excited states unbound. Except for the lowest $2^+$ state, all %the 
$^6$He excited states are either broad resonances or states in the continuum. We observe a good stability of the $2^+_1$ state with respect to the basis size of our NCSM calculations. The higher excited states, however, drop in energy with increasing $N_{\rm max}$ with the most dramatic example being the multi-$\hbar\Omega$ $0^+_3$ state. This spells a potential difficulty for a NCSM/RGM calculations of $^7$He within a $n+^6$He cluster basis as, with increasing density of $^6$He states at low energies, a truncation to just a few lowest eigenstates becomes questionable. 

For the $^7$He, the NCSM predicts the g.s.\ unbound in agreement with experiment. However, the resonance energy with respect to the $^6$He$+n$ threshold appears overestimated. 
Obviously, it is not clear that the ad hoc exponential extrapolation is valid for 
unbound states. 
In addition, no information on the width of the resonance can be obtained from the NCSM calculation. 
\begin{table}[b]
\begin{center}
\begin{ruledtabular}
\begin{tabular}{cc|c|c|cc|c}
$^7$He $J^\pi$  & $^6$He${-}n(lj)$      & NCSM  & CK      & VMC & GFMC & Exp. \\
\hline
$3/2^-_1$         &  $0^+{-}p\frac{3}{2}$ & 0.56     & 0.59   & 0.53   & 0.565 & 0.512(18)~\cite{Cao2012} \\
                         &                                   &             &           &           &           & 0.64(9)~\cite{Beck2007} \\
                         &                                   &             &           &           &           & 0.37(7)~\cite{Wuosmaa_2005} \\
$3/2^-_1$         &$2^+_1{-}p\frac{1}{2}$& 0.001    &  0.06  &  0.006&          &   \\
$3/2^-_1$         &$2^+_1{-}p\frac{3}{2}$& 1.97     &  1.15  &  2.02   &          &   \\
$3/2^-_1$         &$2^+_2{-}p\frac{1}{2}$&  0.12    &        &  0.09      &           &   \\
$3/2^-_1$         &$2^+_2{-}p\frac{3}{2}$&  0.42    &        &  0.30      &           &   \\
$1/2^-$            & $0^+{-}p\frac{1}{2}$&  0.94    & 0.69   &  0.91     &           &   \\
$1/2^-$            &$2^+_1{-}p\frac{3}{2}$&  0.34   & 0.60   &  0.26    &           &   \\
$1/2^-$            &$2^+_2{-}p\frac{3}{2}$&   0.93   &           &            &           &   \\
%$5/2^-$            &$2^+_1{-}p\frac{1}{2}$&   0.77   & 0.85   &  0.81    &          &   \\
%$5/2^-$            &$2^+_1{-}p\frac{3}{2}$&   0.49   & 0.52   &  0.37    &           &   \\
%$5/2^-$            &$2^+_2{-}p\frac{1}{2}$&   0.26   &           &            &           &   \\
%$5/2^-$            &$2^+_2{-}p\frac{3}{2}$&   1.30   &           &            &           &   \\
%$3/2^-_2$         &   $0^+{-}p\frac{3}{2}$&  0.06    &  0.06   &  0.05  &          &   \\
%$3/2^-_2$         &$2^+_1{-}p\frac{1}{2}$&  1.10     &  1.05   &  1.07  &         &   \\
%$3/2^-_2$         &$2^+_1{-}p\frac{3}{2}$&   0.08   &   0.32   &  0.03   &           &   \\
%$3/2^-_2$         &$2^+_2{-}p\frac{1}{2}$&   0.03    &           &            &           &   \\
%$3/2^-_2$         &$2^+_2{-}p\frac{3}{2}$&   0.25    &           &            &           &   \\
\end{tabular}
\end{ruledtabular}
\caption{NCSM spectroscopic factors compared to Cohen-Kurath (CK)~\cite{CK} 
and VMC/GFMC~\cite{GFMC,Brida2011,Wiringa} calculations and experiment. 
The CK values should be still multiplied by $A/(A{-}1)$ to correct for the center of mass motion.}
\label{tab:specfac}
\end{center}
\end{table}
We can, however, study the structure of the $^7$He NCSM eigenstates 
by calculating their overlaps with $^6$He$+n$ cluster states, which are related to $\bar{g}_{\lambda\nu}$ (see Eq.~(\ref{eq:formalism_20})), and the corresponding spectroscopic factors summarized in Table~\ref{tab:specfac}. 
Overall, we find a very good agreement with the VMC/GFMC results as well as with the latest experimental value for the g.s.~\cite{Cao2012}. Interesting features to notice is %the spread of the $3/2^-$ ground state wave function over all three considered $^6$He states with the dominance of the $2^+_1$ and
 the about equal spread of $1/2^-$ between the $0^+$ and $2^+_2$ states. 
We stress that in our present calculations, the overlap functions and spectroscopic factors are not the final products to be compared to experiment but, on the contrary, inputs to more sophisticated NCSMC calculations.     

%\subsection{$^7$He  NCSM/RGM and NCSMC calculations}
%\label{subsec:7He_NCSMC}

The NCSM/RGM calculations for the $n+^6$He system presented in the following were obtained by including up to the three lowest eigenstates of $^6$He, {\em i.e.}, $0^+, 2^+_1$, and  $2^+_2$. These results will be compared to NCSMC calculations, which couple the above $n+^6$He  binary-cluster states to the 6 lowest negative parity NCSM eigenstates of $^7$He ($3/2^-_1,1/2^-,5/2^-,3/2^-_2,3/2^-_3,3/2^-_4$) as well as the four lowest $^7$He positive-parity eigenstates ($1/2^+,5/2^+_1,3/2^+,5/2^+_2$). 

\begin{figure}[t]
\begin{center}
\includegraphics[clip=,width=0.39\textwidth]{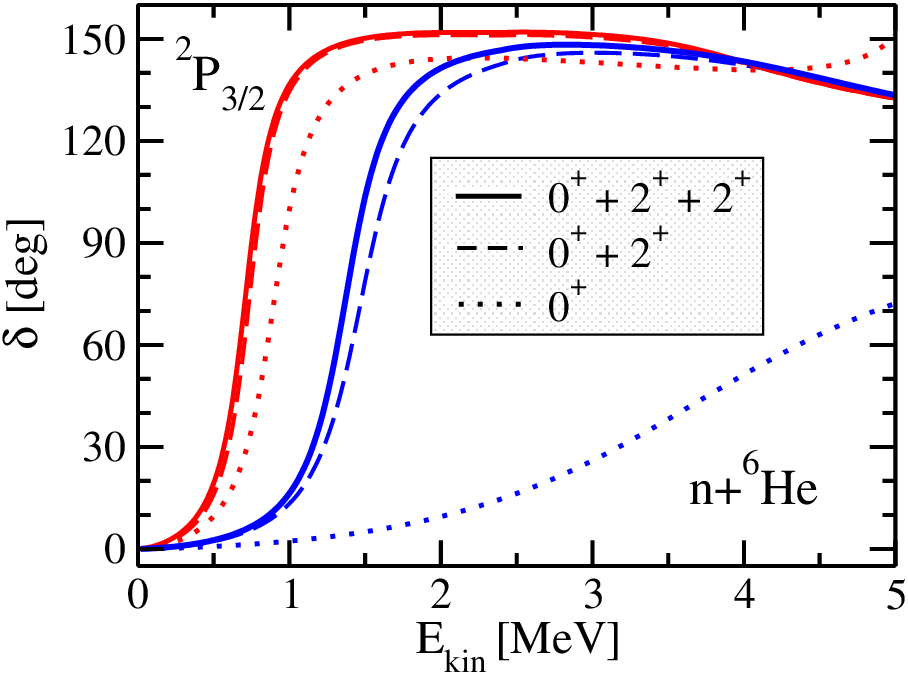}
\end{center}
\caption{(color online). Dependence of the NCSM/RGM (blue lines) and NCSMC (red lines) $\elemA{6}{He} + n$ 
%Sim
diagonal
phase shifts of the $^7$He $3/2^-$ g.s.\
on the number of $^6$He states included in the binary-cluster basis. The short-dashed, 
dashed, and solid curves correspond to calculations
with the $\elemA{6}{He}$ $0^+$ g.s.\ only, $0^+,2^+$ states, and $0^+,2^+,2^+$ states, respectively. 
}
\label{6He_n_target_states}
\end{figure}
First, in Fig.~\ref{6He_n_target_states}, we study the dependence of the $3/2^-$ g.s.\ 
diagonal
phase shifts on the number of $^6$He eigenstates included in the 
NCSM/RGM (blue lines) and NCSMC (red lines) calculations. 
The NCSM/RGM calculation with the $^6$He target restricted to its 
g.s.\ does not produce a $^7$He $3/2^-$ resonance (the phase shift does not reach 90 degrees). A $^2P_{3/2}$ resonance does appear once the $2^+_1$ state of $^6$He is coupled, and the %phase shift is further shifted, the
resonance position further moves to lower energy with the inclusion of the second $2^+$ state of $^6$He. 
On the contrary, the $^2P_{3/2}$ resonance is already present in the NCSMC calculation 
with only the 
g.s.\ of $^6$He. 
In fact, this NCSMC model space is already enough to 
obtain the $\elemA{7}{He}$ $3/2^-$ g.s.\ resonance at about $1 \mev$ above threshold, which is lower than the NCSM/RGM prediction of $1.39 \mev$ when three $\elemA{6}{He}$ states are included. Adding the $2^+_1$ state of $^6$He generates a modest shift of the resonance to a still lower energy while the second $2^+$ state of $^6$He 
has no significant influence (Fig.~\ref{6He_n_target_states}, panel (b)). We further observe that the resonance position in the NCSMC calculation is lower than the NCSM/RGM one by about 0.7 MeV. This difference is due to the additional correlations brought by the $^7$He eigenstates that are coupled to the $n+^6$He binary-cluster states in the NCSMC and that 
compensate for higher excited states of the $^6$He target omitted in the NCSM/RGM sector of the basis. These include both positive-parity states, some of which are shown in Fig.~\ref{6He_exct}, and negative-parity excitations, e.g., the $1^-$ soft dipole excitation {\em etc.} While NCSM/RGM calculations with a large number of clusters' excited states 
can become prohibitively expensive, the coupling of a few NCSM eigenstates of the composite system is straightforward. 

\begin{figure}[t]
\begin{center}
\includegraphics[clip=,width=0.4\textwidth]{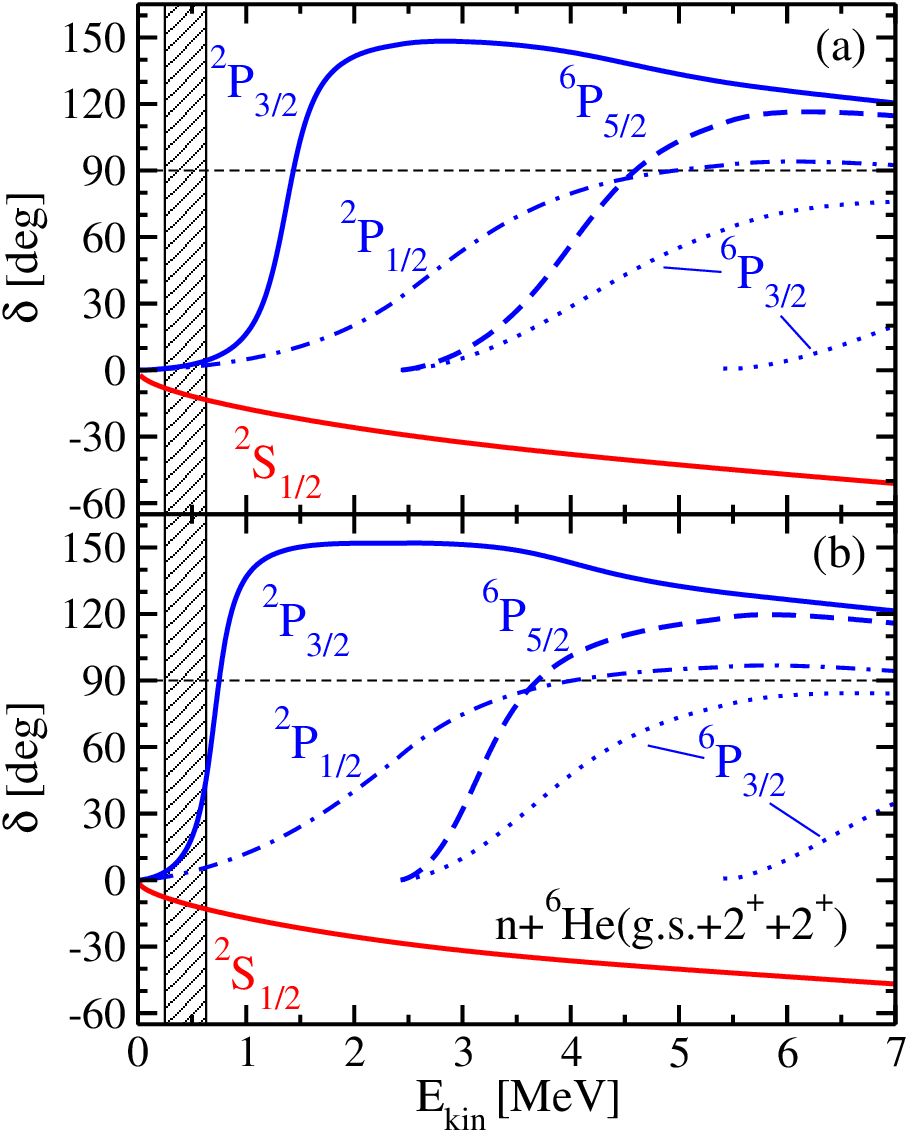}
\end{center}
\caption{(color online). NCSM/RGM (a) and NCSMC (b) $\elemA{6}{He}+n$ diagonal phase shifts (except $^6P_{3/2}$, which are eigenphase shifts)
as a function of the kinetic energy in the center of mass.
The dashed vertical area centered at $0.43 \mev$ indicates the experimental
centroid and width of the $\elemA{7}{He}$ g.s.~\cite{Stokes_1967,Cao2012}.
In all calculations the lowest three $\elemA{6}{He}$ states
have been included in the binary-cluster basis. 
See text for further details.}
\label{6He_n_phase_shifts}
\end{figure}
The NCSM/RGM and the NCSMC phase shifts for the $n+\elemA{6}{He}$ five $P$-wave 
and $^2S_{1/2}$ channels are shown in Fig.~\ref{6He_n_phase_shifts}. All curves 
have been obtained including the lowest three $\elemA{6}{He}$ states.  
The NCSMC calculations (panel (b)) additionally include ten $\elemA{7}{He}$ NCSM eigenstates, as described above.
As expected from a variational calculation, the introduction of the additional $A$-body
basis states $\ket{A \lambda J^\pi T}$ 
lowers 
the centroid values for all $\elemA{7}{He}$ resonances when going from NCSM/RGM (panel (a)) to NCSMC (panel (b)). 
In particular, the $\elemA{7}{He}$ $3/2^-$ g.s.\ 
and $5/2^-$ excited state are pushed toward 
the $\elemA{6}{He}{+}n$ threshold, 
closer to their respective experimental positions. 

The experimental accepted values for the resonance centroids in $\elemA{7}{He}$ and the possible $1/2^-$ states are shown in Table~\ref{6He_n_table_widths}, together with our calculations.  
For NCSM/RGM and NCSMC, the resonance centroids $E_R$ are obtained as the values of the kinetic energy in the center of mass for which the first derivative of the phase shifts is maximal~\cite{Thompson_priv}. 
The resonance widths are then computed from the phase shifts according to (see, e.g., Ref.~\cite{Thompson}) 
\begin{equation}\label{eq:6Hen_10}
  \Gamma=\left. \frac{2}{{\rm d} \delta(E_{kin}) / {\rm d} E_{kin}}\right|_{E_{kin}=E_R}\,.
\end{equation}
An alternative, less general, choice for the resonance energy $E_R$ could be the kinetic energy corresponding to a phase shift of $\pi/2$ (thin dashed lines in Fig.~\ref{6He_n_phase_shifts}). 
While 
Eq.~(\ref{eq:6Hen_10}) 
is safely applicable to sharp resonances, broad resonances
would require an analysis of the scattering matrix in the complex plane. 
As we are more interested in a qualitative discussion of the results,
we use here the above extraction procedure for broad resonances as well.
The two alternative ways of 
choosing $E_R$ 
lead to basically identical results for the calculated $3/2^-_1$ resonances, however the same is not true for the broader $5/2^-$ %resonance 
and the very broad $1/2^-$ resonances. 
The 
$\pi/2$ condition, 
particularly questionable for broad resonances, would result in $E_R\sim 3.7$ MeV and $\Gamma\sim 2.4$ MeV for the $5/2^-$  and $E_R\sim 4$ MeV (see Fig.~\ref{6He_n_phase_shifts}) and $\Gamma\sim 13$ MeV for the $1/2^-$ resonance, respectively. 
\begin{table}[t]
\begin{center}
\begin{ruledtabular}
\begin{tabular}{c|ccc|cc|cc|c}
$J^\pi$        & \multicolumn{3}{c|}{experiment} & \multicolumn{2}{c|}{NCSMC} & \multicolumn{2}{c|}{NCSM/RGM} & NCSM \\ 
               &  $E_R$   & $\Gamma$ & Ref.                  &   $E_R$  & $\Gamma$   &  $E_R$  & $\Gamma$ & $E_R$    \\
\hline
$3/2^-$        & 0.430(3)& 0.182(5)& \cite{Cao2012}     &  0.71 & 0.30  & 1.39  & 0.46       & 1.30 \\
$5/2^-$        & 3.35(10) & 1.99(17) & \cite{Tilley2002} &  3.13 & 1.07 & 4.00  & 1.75        & 4.56 \\
$1/2^-$        & 3.03(10) & 2          & \cite{Wuosmaa_2005}   &  2.39 & 2.89 & 2.66  & 3.02        & 3.26 \\
                     & 3.53       & 10         & \cite{Boutachkov_2005} &   &   &       &   &   \\
                     & 1.0(1)     & 0.75(8)  & \cite{Meister_2002} &   &   &       &   &   \\
\end{tabular} 
\caption{Experimental and theoretical %values for the 
resonance centroids and widths in MeV for the
$3/2^-$ g.s.\ , $5/2^-$ and $1/2^-$ excited states of $\elemA{7}{He}$. 
See the text for more details.}
\label{6He_n_table_widths}
\end{ruledtabular}
\end{center}
\end{table}

The resonance position and width of our NCSMC $3/2^-$ g.s.\  slightly overestimate the measurements, 
whereas the prediction for the $5/2^-$ 
is lower compared to experiment~\cite{Korsheninnikov_1998,Tilley2002}, although our determination of the width should be taken with some caution in this case.
As for the $1/2^-$ resonance, the experimental situation is not clear as discussed in the introduction and documented in Table~\ref{6He_n_table_widths}. While the centroid energies 
of Refs.~\cite{Wuosmaa_2005,Wuosmaa_2008}  and~\cite{Boutachkov_2005} are  comparable, the widths are very different. With our determination of $E_R$ and $\Gamma$, the NCSMC results are in fair agreement with the neutron pick-up and proton-removal reactions experiments~\cite{Wuosmaa_2005, Wuosmaa_2008} 
and definitely do not support the hypothesis of a low lying ($E_R{\sim} 1$ MeV) narrow ($\Gamma \leq 1$ MeV) $1/2^-$ resonance~\cite{Markenroth_2001,Meister_2002,Skaza_2006,Ryezayeva_2006,Lapoux_2006}.
In addition, our NCSMC calculations predict two broad $^6P_{3/2}$ resonances (from the coupling to the two respective $^6$He $2^+$ states) at about 3.7 MeV and 6.5 MeV with widths of 2.8 and 4.3 MeV, respectively. The corresponding eigenphase shifts do not reach $\pi/2$, see Fig.~\ref{6He_n_phase_shifts}. In experiment, there is a resonance of undetermined spin and parity at 6.2(3) MeV with a width of 4(1) MeV~\cite{Tilley2002}.
Finally, it should be noted that our calculated NCSMC ground state resonance energy, 0.71 MeV, is lower but still compatible with the extrapolated NCSM value of 0.98(29) MeV (see Tables~\ref{tab:NCSM_He_gs} and \ref{6He_n_table_widths}).

In conclusion, we introduced a new unified approach to nuclear bound and continuum states based on the coupling of the no-core shell model with the no-core shell model/resonating group method. We demonstrated the potential of the NCSMC in calculations of $^7$He resonances. Our calculations do not support the hypothesis of a low lying $1/2^-$ resonance in $^7$He. 

%The advantages of the NCSMC are expected to become even more evident in calculations with composite projectiles (such as deuteron, $^3$H, or $^3$He) that require the use of a large number of pseudostates in the NCSM/RGM to account for virtual breakup effects. The contribution of the pseudostates is expected to be suppressed in the NCSMC approach. Extension of the NCSMC formalism to the case of composite projectiles, the inclusion of three-nucleon interactions, and the coupling of three-body clusters are under way.

\acknowledgments

Computing support came in part from the LLNL institutional Computing Grand Challenge program. Prepared in part by LLNL under Contract DE-AC52-07NA27344. Support from the U.\ S.\ DOE/SC/NP (Work Proposal No.\ SCW1158) and the Natural Sciences and Engineering Research Council of Canada (NSERC) Grant No.\ 401945-2011 is acknowledged. TRIUMF receives funding via a contribution through the National Research Council Canada. This research was supported in part by the PAI-P6-23 of the Belgian Office for Scientific Policy and by the European Union Seventh Framework Programme under grant agreement No.\ 62010.

\bibliographystyle{apsrev}

\end{document}